\documentclass{webofc}
\usepackage[varg]{txfonts}   
\begin{document}
\title{Deciphering the phases of QCD matter with fluctuations and correlations of conserved charges}
%
%

\author{\firstname{Anar} \lastname{Rustamov}\inst{1,2}\fnsep\thanks{\email{a.rustamov@gsi.de}} 
}

\institute{GSI Helmholtzzentrum f{\"u}r Schwerionenforschung, Darmstadt, Germany
\and
           National Nuclear Research Center, Baku, Azerbaijan 
          }

\abstract{%
A review is given on recent experimental and theoretical/phenomenological developments regarding the phase structure of the strongly interacting matter. Specifically, evolution with the collision energy of net-proton number fluctuations as measured by several experiments are presented and their implications for the QCD phase diagram are outlined. In addition, theoretical calculations on correlations between conserved charges are presented and prospects for their experimental explorations are addressed. 
}
\maketitle
\section{Introduction}
\label{intro}
The quest for the phase structure of strongly interacting matter, hereinafter referred to as Quantum Chromodynamics (QCD) matter, remains at the focus of contemporary theoretical and experimental investigations.  By know it is well established that, in nuclear collisions at cm energies from several TeV down to about 12 GeV per nucleon pair the matter created in head-on collisions of heavy ions freezes out close to the chiral phase transition line, taking place at a pseudo-critical temperature values of about 156 MeV~\cite{Andronic:2017pug, HotQCD:2018pds}. This opens unique experimental opportunities to probe the strongly interacting matter close to the phase boundary, such as experimental verification of its Equation of State (EoS). In this energy range the lattice QCD (lQCD) calculations predict crossover chiral transition~\cite{HotQCD:2018pds}, which still awaits experimental confirmations. Contrary to the high energy region, first principle calculations at low energies (high net-baryon densities) are still not available, however a number of effective models predict, that at reasonably large net-baryon densities the QCD matter undergoes a first order chiral phase transition~\cite{Asakawa:1989bq, Rajagopal:1992qz, Stephanov:1998dy, Stephanov:2006dn, Sasaki:2007db}. This lends support to the existence of the chiral Critical End Point (CEP) at which the anticipated first order phase transition line terminates and smooth crossover transition sets in. Hence at the CEP the system would undergo a second order phase transition. One of the main goals of experiments at lower energies is to locate the position of the chiral CEP in the QCD phase diagram. 

Fluctuations of conserved charges, such as baryon number, electric charge, strangeness and charm, as well as correlations between them are identified as promising tools to probe the chiral criticality. The objective is to characterise the response of the system to external perturbations, such as infinitesimal changes in the chemical potentials. In theoretical calculations, within the Grand Canonical Ensemble (GCE) of statistical mechanics, for a system of fixed volume $V$ and temperature $T$ this task is performed by evaluating the $n^{th}$ order susceptibilities $\chi_{q}^{n}$, i.e.,  derivatives of the thermodynamic pressure $P$ with respect to the chemical potentials $\mu_{q}$ responsible for the conservation of the corresponding charge $q$ on average, evaluated at vanishing values of chemical potentials

\begin{equation}
\label{eq-chi}
    \hat{\chi}_{n}^q= \left.\frac{\partial^{n}\hat{P}}{\partial \hat{\mu}_{q}^n}\right\rvert_{\vec{\mu} = 0}  = \frac{\kappa_{n}(N_{q})}{VT^3},
\end{equation}
where $\hat{\chi}_{n}^q$, $\hat{P}=P(T, V, \vec{\mu})/T^{4}$ and $\hat{\mu}_{q}=\mu_{q}/T$ denote reduced susceptibility, reduced pressure and reduced chemical potential, respectively. 
In experiments the cumulants are directly measured in a given detector acceptance either by first reconstructing the multiplicity distributions of the corresponding net-charge number ($N_{q}$) or by using probabilistic approaches based on the detector response functions for particle identification (cf.~\cite{Rustamov:2012bx, Arslandok:2018pcu} and references therein). The so obtained experimental cumulants $\kappa_{n}(N_{q})$ can be directly compared to those computed via derivatives of thermodynamic pressure (cf. Eq.~\ref{eq-chi}), provided that the system under consideration is in thermal equilibrium and all necessary instrumental effects and non-critical signals, such as those stemming from detection efficiencies, participant fluctuations, suppression of fluctuations due to e-by-e conservation laws and resonance decays are properly accounted for~\cite{Nahrgang:2009dqc, Braun-Munzinger:2016yjz, Rustamov:2020ekv, Braun-Munzinger:2020jbk}.

\section{Experimental Results}
In this section I briefly discuss experimental measurements related to crossover and critical point studies, provide their comparison with the model calculations and  implications for the QCD phase diagram.

\subsection{Search for a critical point}
\begin{figure}[h]
\centering
\includegraphics[width=13cm,clip]{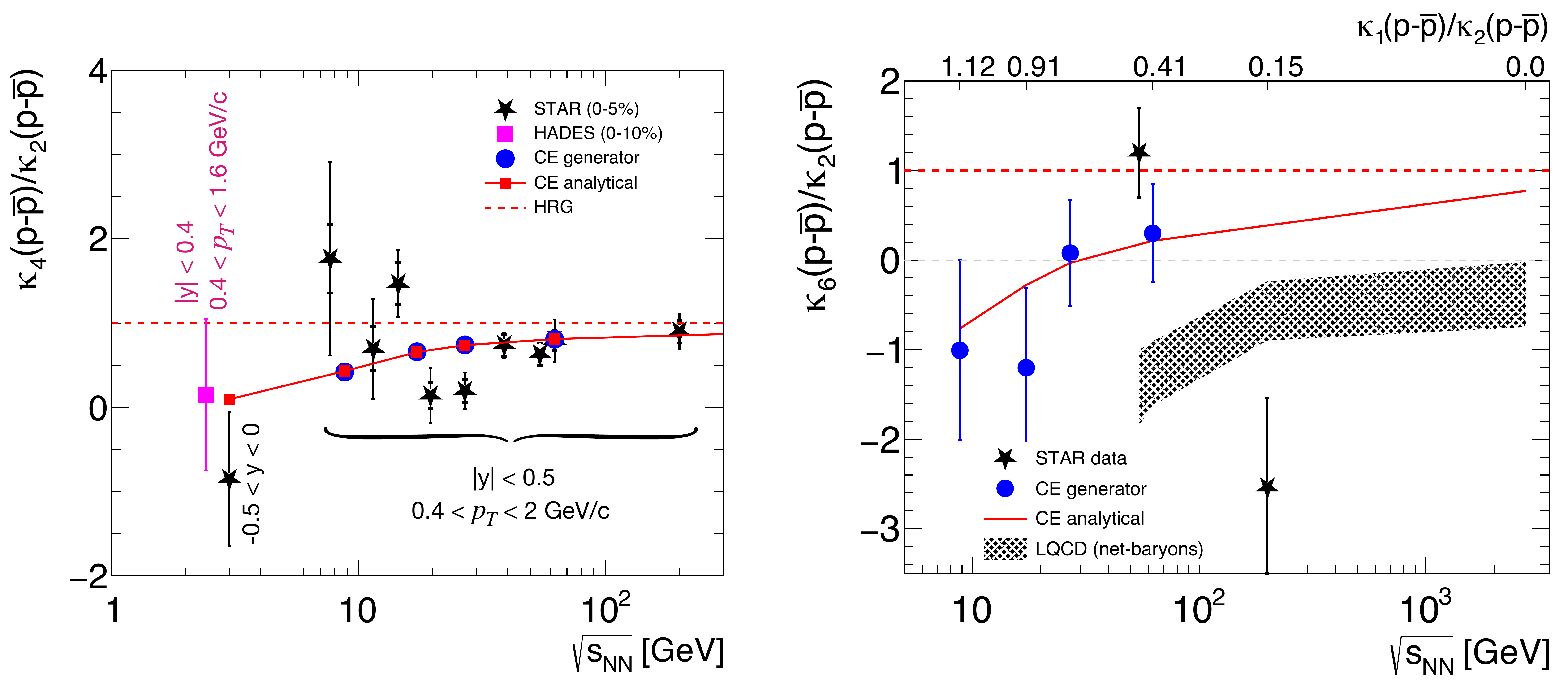}
\caption{Left panel: evolution with the collision energy of $\kappa_{4}/\kappa_{2}$ measured for net-proton distributions. The STAR data, represented with black star symbols, are measured for |y|<0.5 (-0.5 <y < 0 for the lowest energy) and 0.4 < p$_{T}$[GeV/c] < 2. The HADES measurement (magenta square symbol) is for |y| < 0.4 and 0.4 < p$_{T}$[GeV/c] < 1.6. Also presented are two non-critical baselines; the HRG baseline: indicated with the dashed line at unity, the CE baseline: shown by solid the red line/boxes (analytical calculations) and blue symbols (generated values). Right panel: Similar to the left panel for $\kappa_{6}/\kappa_{2}$. The corresponding lQCD results for net-baryons are presented with the hashed region.}
\label{fig-k4} 
\end{figure}
Fig.~\ref{fig-k4} shows the evolution with collision energy of the fourth to seconder order cumulants of net-proton distributions\footnote{At low collisions energies these are essentially cumulants of proton number.} as measured by the STAR~\cite{STAR:2020tga, STAR:2021fge} (black star symbols) and HADES~\cite{HADES:2020wpc} (magenta box) collaborations. The data points are contrasted with calculations within two different ensembles of statistical mechanics~\cite{Braun-Munzinger:2018yru, Braun-Munzinger:2019yxj, Braun-Munzinger:2020jbk}: (i) Grand Canonical Ensemble (GCE) with the ideal gas EoS in the Boltzmann limit, represented with the dashed line along unity; (ii) Canonical Ensemble (CE), implemented in the full phase space, also using the ideal gas EoS. In the following these are referred to as the GCE or Hadron Resonance Gas (HRG) baseline and the CE baseline, respectively. Such treatment of cumulants within CE ensures that in each event the baryon number (number of baryons minus number of antibaryons) is exactly conserved in the full phase space, inducing strict correlations between baryon and anti-baryon numbers. As a consequence, since there are essentially no antibaryons at low energies, the baryon number does not fluctuates in the full phase space, leading to a binomial distributions in the finite acceptance at low energies (for more details see the Appendix C of Ref.~\cite{Braun-Munzinger:2020jbk})\footnote{At high energies baryons and anti-baryons fluctuate in the full phase space, only net-baryon number does not.}. This, however, apples if wounded nucleons~\cite{Bialas:1976ed} do not fluctuate from event-to-event, while in practice they do~\cite{Braun-Munzinger:2016yjz, Skokov:2012ds}. Thus, non-critical contributions to the finally measured (net-)proton cumulants, stemming from fluctuations of wounded nucleons, need to be corrected for. This is in particular essential at low energies because, in addition to the above mentioned arguments regarding binomial multiplicity distributions, most of these contributions scale with the net-proton number. For example, at LHC energies, where net-proton number at mid-rapidity is practically zero, contributions from wounded nucleon fluctuations are immaterial for the measured second order cumulants of net-protons, while at lower energies these contributions are substantial~\cite{Braun-Munzinger:2016yjz, HADES:2020wpc, STAR:2021fge}. Coming back to Fig.~\ref{fig-k4}, both analytic (red line/boxes) and generated (blue symbols) results of the CE baseline are presented there. It is worthwhile to mention that, ingredients of the CE calculations, such as single particle partition functions are estimated using the experimentally measured mean multiplicities of protons, which are "naturally" treated as CE multiplicities. Also, it is important to note that the analytic calculations and generated values fully incorporate experimental conditions, such as acceptances both in rapidity and transverse momentum. Fig~\ref{fig-k4} shows that correlations induced by the baryon number conservation in full phase space survive in the finite experimental acceptance thereby suppressing, with respect to the GCE baseline, the net-proton cumulants. Clearly this kind of modifications of cumulants do not originate from searched for critical phenomena induced by phase transitions or existence of a critical point. The CE results represent more realistic ("proper") baseline. The critical behaviour due the presence of the critical point is predicted based on universality arguments and suggest non-monotonic behaviour of the order  parameter as a function of collision energy, which should led to non-monotonic, with respect to a "proper" baseline, energy dependence of fourth order cumulants for proton or net-proton number distributions~\cite{Stephanov:2011pb}. Such a non-monotonic behaviour with the significance of 3.1$\sigma$ is reported by the STAR collaboration~\cite{STAR:2020tga}, albeit with respect to the GCE baseline. However, this conclusion changes if one considers the "proper" baseline. Indeed, within the uncertainties, the experimentally measured energy dependence of the $\kappa_{4}/\kappa_{2}$ ratio, presented in the left panel of Fig.~\ref{fig-k4}, are consistent with the CE, i.e., noncritical baseline, as reported in Ref.~\cite{Braun-Munzinger:2020jbk}. The presented experimental measurements look promising, however the  precision of the data does not allow to draw firm conclusions regarding the existence of the first order phase transition or related CEP. This situation will improve significantly with the available high statistics data from the STAR experiment, the ongoing analysis at NA61/SHINE and future measurements at the CBM experiment~\cite{Friman:2011zz}.

\subsection{Search for a crossover transition}
The search for the crossover transition has a clear advantage because of its closeness to the genuine second order phase transition belonging to the $O$(4) universality class in three dimensions, realised in the limit of massless u and d quarks~\cite{Pisarski:1983ms}. Indeed, lQCD calculations indicate that critical signals of the chiral limit survive also after accounting for physical light quark masses~\cite{Ejiri:2005wq}. On the other hand, the cumulants of the $O$(4) dynamics significantly differ from those calculated within GCE with the ideal gas EoS~\cite{Friman:2011pf}. For example, the sixth (also eighth) order cumulants of net-baryons are negative, while in GCE with the ideal gas EoS they remain positive such that $\kappa_{6}/\kappa_{2}=\kappa_{8}/\kappa_{2}=1$. Here one again needs to be careful, because the CE baseline for $\kappa_{6}/\kappa_{2}$  may also become negative starting from some collision energy~\cite{Braun-Munzinger:2020jbk, Vovchenko:2021kxx}. Hence deviations of these ratios from the CE baseline towards negative values would be a strong signal for the crossover transition~\cite{Almasi:2017bhq, Fu:2021oaw, Borsanyi:2018grb, Bazavov:2020bjn}, still to be confirmed in experiments. In the right panel of Fig.~\ref{fig-k4}  the ratios of the six to second order cumulants of net-protons are presented, as measured by the STAR experiment. Qualitatively the 200 GeV STAR data are indeed consistent with the anticipated negative sign.  However, the data point at  54.4 GeV of STAR is at odds with the lQCD calculations~\cite{Bazavov:2020bjn}. Another interesting measurements would be to test the hierarchy of cumulant ratios as predicted by the lQCD calculations for baryon number susceptibilities. Such measurements were recently reported by the STAR collaboration for net-protons~\cite{STAR:2022vlo}, yet with sizable measurement uncertainties. Future as well as already available data from the STAR and ALICE collaborations will allow for systematic exploration of the crossover region of the QCD phase diagram with significantly enlarged statistics. 
\begin{figure}[h]
\centering
\includegraphics[width=13cm,clip]{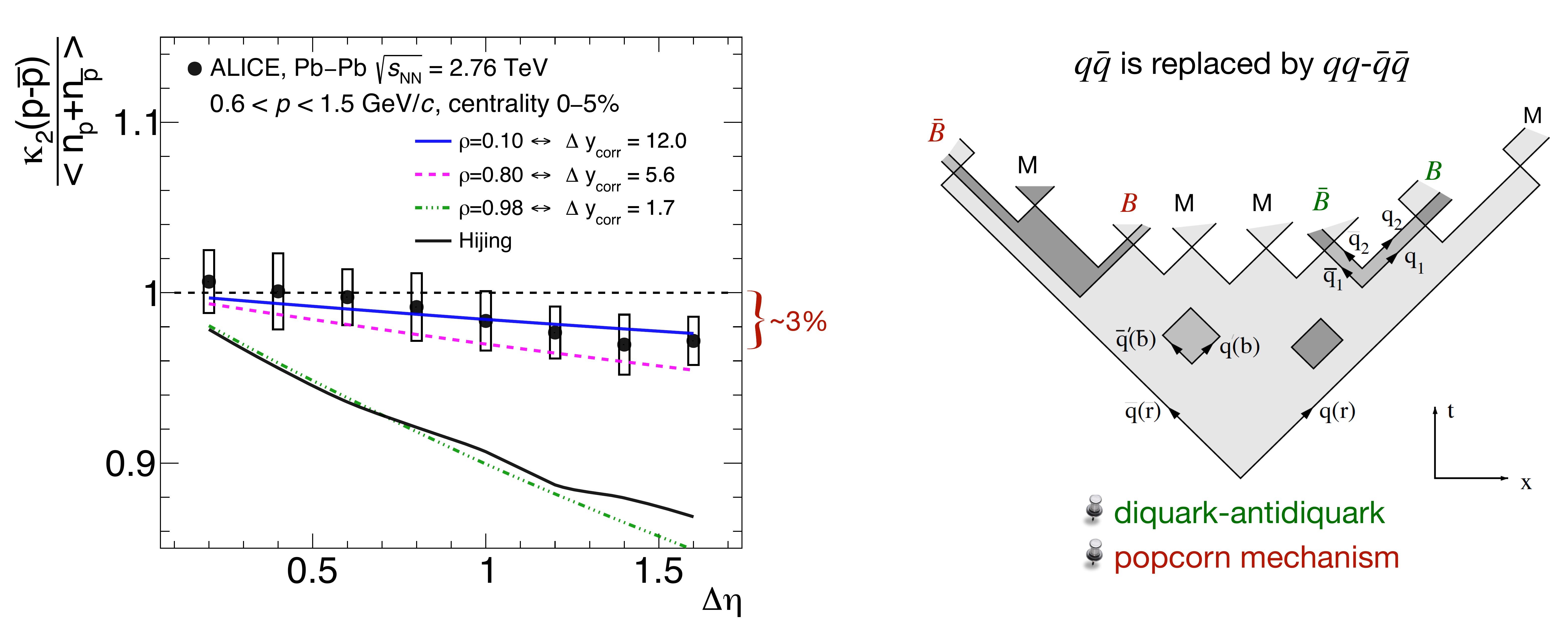}
\caption{Left panel: scaled second order cumulants of net-proton distributions as a function of the acceptance coverage around mid-rapidity, measured by the ALICE experiment (black symbols)~\cite{Rustamov:2017lio, ALICE:2019nbs}. The colored lines correspond to calculations accounting for baryon number conservation for different values of the correlation coefficient in the rapidity space~\cite{Braun-Munzinger:2019yxj, LocalBaryonQM22}. The solid black line represents the results from the HIJING event generator. Right panel: a schematic illustration of baryon production mechanisms via the Lund string model.}
\label{fig-k2-ALICE} 
\end{figure}

The acceptance dependence of the normalized second order cumulants of net-proton distributions as measured by the  ALICE experiment are presented in the left panel of Fig.~\ref{fig-k2-ALICE}~\cite{Rustamov:2017lio, ALICE:2019nbs}. The ALICE data are compared to both GCE and CE baselines~\cite{Braun-Munzinger:2018yru, Braun-Munzinger:2020jbk}. Furthermore, the CE baseline is performed for different correlations between baryons and anti-baryons quantified with the Pearson correlation coefficient $\rho$ and/or the correlation length $\Delta y_{corr}$~\cite{Braun-Munzinger:2019yxj, LocalBaryonQM22}. In doing so either Cholesky factorisation~\cite{Cholesky} or the Metropolis algorithm~\cite{Metropolis:1953am} is used. The $\rho=0$ case corresponds to the global baryon number conservation or long range correlations.  As seen from Fig.~\ref{fig-k2-ALICE} the experimental measurements are best described with $\rho = 0.1$, i.e., the data suggest long range correlations between baryons and antibaryons in the rapidity space. Thus, after accounting for the global baryon number conservation the ALICE data on second order cumulants of net-protons is consistent with the corresponding cumulants of the Skellam distribution. This is in line with the lQCD predictions of a Skellam behaviour for the second order cumulants of net-baryon distributions at a pseudo-critical temperature of about 156 MeV. The results from the HIJING event generator, shown with the black solid line in Fig.~\ref{fig-k2-ALICE}, grossly underestimate experimental measurements. The large discrepancy between the ALICE data and HIJING simulations calls into question the baryon production mechanism used in the Lund string  model, namely by producing them via diquark-antidiquark pairs~\cite{Andersson:1983ia}, schematically illustrated in the right panel of Fig.~\ref{fig-k2-ALICE}. The latter leads to a small correlation lengths in the rapidity space, which in turn, as presented in Fig.~\ref{fig-k2-ALICE}, suppresses fluctuations of net-baryons; but much stronger compared to experimental observations. Indeed, to catch the trend of the HIJING results one needs to introduce rather strong correlation coefficient of about $\rho$=0.98 (cf. Fig.~\ref{fig-k2-ALICE}). Such short-range rapidity correlations from models can be softened by exploiting popcorn mechanism for baryon production, in which additional meson (even several mesons) is produced in between the diquark-antidiquark pairs~\cite{Eden:1996xi}. The presented ALICE measurements on net-proton fluctuations as a function of the rapidity coverage demonstrate that baryon production at high energies is not fully understood and are useful to discriminate between several baryon production mechanisms available in the Lund string model.   

\begin{figure}[h]
\centering
\includegraphics[width=13cm,clip]{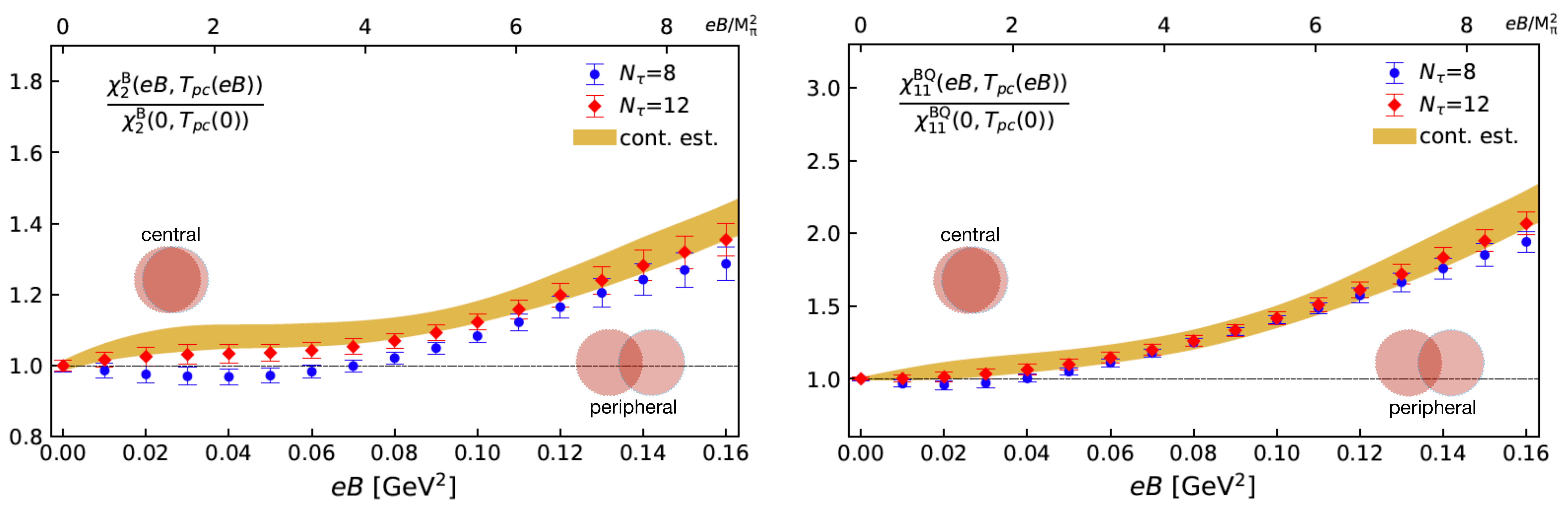}
\caption{Left panel: normalised second order fluctuations of baryon number along the transition line as a function of the magnetic field strength. Right panel: similar to the left panel for the normalized second order correlations between baryon number and electric charge.}
\label{fig-k2-Magnet} 
\end{figure}

\subsection{Fluctuations in the presence of a magnetic field}
Fluctuations of conserved charges can serve as ideal probes of a magnetic field in nuclear collisions~\cite{Kharzeev:2007jp, Skokov:2009qp}. First (2+1)-flavor lQCD calculations, with physical quark masses, on the second order susceptibilities for baryon number and correlations between baryon number and electric charge as a function of a magnetic field strength $eB$ are presented in Fig.~\ref{fig-k2-Magnet}, where fluctuations without magnetic field are used as a reference~\cite{Ding:2021cwv, Ding:2022uwj}. As seen from Fig.~\ref{fig-k2-Magnet}, from central towards peripheral collisions the ratio of susceptibilities progressively deviate from unity, induced by the presence of a magnetic field. The correlations between different conserved charges (right panel of Fig.~\ref{fig-k2-Magnet}) exhibit even more pronounced sensitivity to the presence of the magnetic field. In experiments the signal can be looked at by measuring net-baryon cumulants or mixed cumulants between different conserved charges as a function of centrality normalized to the results obtained for central collisions. 

\section{Acknowledgments}
The author acknowledges stimulating discussions with Mesut Arslandok, Peter Braun-Munzinger, Bengt Friman, Romain Holzmann, Krzysztof Redlich, Johanna Stachel, Joachim Stroth and Nu Xu.

\bibliography{bibliography}
\end{document}